\documentclass[preprint]{raa}            

\usepackage{graphicx,times,natbib,url}             

\bibpunct{(}{)}{;}{a}{}{,}

\usepackage{longtable}

\begin{document}

   \title{The ROSAT Bright Source 1RXS J201607.0$+$251645: An Active
   Algol-type Binary
}

   \volnopage{Vol.9 (2009) No.9, 1035--1048}      
   \setcounter{page}{1}          

   \author{Hua-Li Li
      \inst{1,2,3}
   \and Yuan-Gui Yang
      \inst{1,4}
   \and Wei Su
      \inst{1,2}
   \and Hui-Juan Wang
      \inst{1,2}
   \and Jian-Yan Wei
      \inst{1}
   }

   \institute{National Astronomical Observatories, Chinese Academy of Sciences,
             Beijing 100012, China; {\it lhl@bao.ac.cn}\\
        \and Urumqi Observatory, Urumqi 830011;
        \and
             Graduate University of Chinese Academy of Sciences, Beijing 100049;\\
        \and
             School of Physics and Electric Information, Huaibei Coal Industry Teachers
             College, Huaibei 235000, Anhui Province, China\\
   }
   \date{Received~~2009 Mar.28; accepted~~2009 May 18}

\abstract{ 1RXS J201607.0+251645 is identified to be an eclipsing
binary. We present the preliminary observations in $V$ band with the
0.6-m telescope  for three years and the extensive observations in
$V$ and $R$ band with the 0.8-m telescope for six nights,
respectively. The light curve of the system is $EB$ type. Five light
minimum times were obtained and the orbital period of
$0\fd388058(\pm0\fd00044)$ is determined. The photometric solution
given by the 2003-version Wilson-Devinney program suggests the
binary is a semi-detached system with the photometric mass ratio
$0.895(\pm0.006)$, which probably comprises a G5 primary and an
oversize K5 secondary. The less massive component has completely
filled its Roche lobe, while the other almost fills its Roche lobe
with the filling factor of $93.4\%$. The system shows a varying
O'Connell effect in its phase folded diagrams from 2005 to 2007, and
is X-ray luminous with $\textrm{log} L_{X}/L_{bol} \sim -3.27$.
Possible mechanisms to account for these two phenomena are
discussed. Finally, we infer the binary may be in thermal
oscillation or may evolve into a contact binary.
 }

   \authorrunning{H.-L. Li,  Y.-G. Yang, W. Su, H.-J. Wang, J.-Y. Wei}            
   \titlerunning{The ROSAT Bright Source 1RXS J201607.0+251645: An Active Algol-type Binary}  

   \maketitle

\section{Introduction}           
\label{sect:intro}

Algol type binaries are eclipsing variables with the less massive
component filling its Roche lobe and transferring material to its
companion \citep{Ziolkowski69, Proper89}. They are X-ray emitters
\citep{Shaw96}, similar to solar analogues such as RS CVn binaries
\citep{Umana91}. The driving mechanisms of X-ray emission in these
systems remain unsolved though are generally thought to be related to large
scale mass transfer \citep{Sarna98}, chromospheric activity on the
components \citep{Hall89, Retter05}, or hot disk \citep{Blondin95}.
Finding more of them may help to shed light on their X-ray nature.

The optical variability campaign for strong X-ray sources with the
0.6-m telescope at the Xinglong observatory was launched in 2005,
aiming to study long term chromospheric activity of young stars.
Containing about 200 objects (much have never been identified) in
the sample selected from the ROSAT bright source catalog (RBS)
\citep{Voges99}, the campaign is an excellent resource for finding
new variables. 1RXS J201607.0+251645 ($\alpha_{2000.0}$ =$
20^h16^m07.^s0$,$\delta_{2000.0}$ = $25^\circ16'45.''0$) was
included in our observing program and caused our interest by its
obvious eclipsing effect.

1$\sigma$ positional error for 1RXS J201607.0+251645 is 9$''$,
including 8$''$ systematic error. Within this error circle, we found
a bright star HD 339946 (also known as GSC 02159-01213, TYC
2159-1213-1, and 2MASS J20160698+2516536) lying 8.8 $''$ from the
X-ray position. Thus HD 339946 is taken as the optical counterpart
of 1RXS J201607.0+251645. It is listed in SIMBAD as a G5 star, with
$V$ band magnitude of $10\fm6$. No previous variability description
could be found in General Catalogue of Variable Stars (GCVS)
\citep{Kholopov98} and other literature.

\section{Observations and the orbital period}
\label{sect:Obs}

Our first set of photometric observations of 1RXS J201607.0+251645
was performed with
the 0.6-m telescope at the Xinglong observatory from Nov. 2005 to
Sep. 2007. Typical observing frequency is several times per night under
auto-observation mode. A PI 1024$\times$1024 CCD and Standard
Johnson V filter were used. The field of view is $17'\times17'$
\citep{Xing06}.

Preliminary analysis had revealed the eclipsing effect of the target.
In order to investigate its eclipsing nature in more detail, extensive
observations were conducted in Nov. 2008 with the 0.8-m Tsinghua-NAOC telescope
(TNT) at the Xinglong observatory. TNT employs a PI
1340$\times$1300 CCD, giving the field of view of $11'\times11'$
\citep{Zheng08}. Standard Johnson $V$ band and $R$ band light curves
were obtained. Finding chart made from a $V-$band CCD image taken
by the 0.8-m telescope is shown in Figure 1, with the comparison
star and the check star marked as C1 and C2 (Table 1), respectively.

All photometric reductions were performed with IRAF
\footnotemark{}\footnotetext{IRAF is distributed by the National
Optical Astronomy Observatory, which is operated by the Association
of Universities for Research in Astronomy, Inc., under cooperative
agreement with the National Science Foundation.} including bias and
dark subtraction and flat-field correction. No differential
extinction was applied since the angular distances of all the stars
were small. The resulting photometric accuracy is estimated to be
$0\fm01$ for observations with the 0.6-m telescope and $0\fm005$ for
those with the 0.8-m telescope.

For preliminary observations, phase folded diagrams in Figure 2 were
plotted with a period of $0\fd388$ obtained with the phase
dispersion minimization (PDM) method \citep{Stellingwerf78}.

In 2008, we obtained $648$ and $645$ extensive observations in $V$ and
$R$ bands, which are shown in Figure 3 and listed in Appendix 1. The
light curves vary continuously between eclipses, indicating that the
system is an $EB$ type eclipsing binary \citep{Sterken05}. The
amplitudes of various light are $0\fm52$ and $0\fm51$ in $V$ and $R$
bands, respectively. The primary eclipse is deeper than the
secondary eclipse by up to $0\fm32$ in $V$ band and $0\fm29$ in $R$
band.

Five light minimum times were obtained in the extensive observations (Table 2).
Using the linear least-square method, the ephemeris formula could be
given as follows:
\begin{equation}
Min.I.=\textrm{HJD}2454771.6336(\pm0.0037)+
0.388058(\pm0.00044)\times E.
\end{equation}

\section{Photometric solution}

The photometric solution was made with the 2003 version of
Wilson-Devinney program \citep{Wilson71, Wilson90, Wilson94}. Before
running the DC program, a series of input parameters were fitted.
The temperature of the primary component was initially set at
$5160\mathrm{K}$ \citep{Cox2000}, in accordance with the G5 dwarf
spectral type given in SIMBAD (operated at CDS). The
gravity-darkening coefficients $g_1$ and $g_2$ were set with the
same value of 0.32 \citep{Lucy67}, assuming convective atmospheres
of the two components. The bolometric albedos were fixed as $A_1 =
A_2 = 0.5$ \citep{Rucinski73}. Linear limb-darkening coefficients
for Star 1 were $x_{1V}=0.65$ and $x_{1R}=0.51$, respectively
\citep{Al-Naimiy78}. Meanwhile the limb-darkening coefficients for
Star 2 (i.e., $x_{2V}$ and $x_{2R}$) were determined according to
its temperature $T_2$. Based on the temperatures, convective
atmospheres were assumed.  The commonly adjustable parameters
employed are the orbital inclination $i$, the mass ratio $q$, the
mean temperature of Star 2 $T_{2}$, the potential of the components
$\Omega_{1}$, the monochromatic luminosity of Star 1 $L_{1}$. The
reflection effect was computed with the detailed model of Wilson
(1990). The relative brightness of Star 2 was calculated by the
stellar atmosphere model.

To find a photometric mass ratio, solutions were obtained for a
series of trial values of the mass ratio ($q=0.2-2.0$). For each
value of the mass ratio, the calculation started at mode 2 (i.e.,
detached mode), but the solution always converged to mode 5 (i.e.,
semi-detached mode), suggesting the secondary has completely filled
its Roche lobe. Corresponding mass ratios and squared residuals
are plotted in Fig.4, where the smallest value of
$\Sigma(O-C)^{2}_{i}$ is locating at q $\sim$ 0.8. At this point, the
set of adjustable parameters was expanded including the mass ratio
$q$. After a few trials and correction of mass ratio, a best-fit
solution was achieved at $q=0.895(\pm0.006)$. The photometric
elements are listed in Table 3. Synthesis light curves as solid
lines are plotted in Figure 5, where the quality of the fits is
fairly well in the $V$ and $R$ bands.

\section{Discussions}
\label{sect:discussion}

\subsection{Optical variability outside eclipse}

As shown in Fig.2, earlier observations had revealed the existence
of O'Connell effect
 \citep{Milone68} in 1RXS J201607.0+251645, which was varying along
 the time. To carefully examine how this asymmetry
varied, we plot the primary maximum brightness ($Max. I$, in magnitude), the secondary
maximum brightness ($Max. II$, in magnitude)
and the values of $Max. II - Max. I$ of each phase folded diagram, as shown in Fig. 6.
Clearly, the O'Connell effect was varying dramatically in the end of 2005:
$Max. I$ was brightening by about $0\fm03$ in Nov. 2005 and $Max. II$ was darkening by about
$0\fm04$ in Dec. 2005. Since 2006, the change of asymmetry became mild.

The varying O'Connell effect is also reported in other active binaries,
such as GR Tauri \citep{Gu04}, U Pegasi \citep{Djurasevic01}, XY UMa \citep{Pribulla01}
and EQ Tau \citep{Yuan07},
in which the varying O'Connell effect is usually modeled and explained by
the development and migration of hot spots \citep{Gu04} or dark spots
\citep{Yuan07, Djurasevic01, Pribulla01}.

So we suggest that either chromospheric activity
\citep{Strassmeier89, Strassmeier92, Strassmeier08}
or the impact of transferred mass
stream onto the primary star \citep{Pereira06, Ibanoglu06} is very strong in our target.
Also, the actual situation is likely to be the combination of both processes.

\subsection{The X-ray emission}

Based on the conversion relation given by Schmitt (1995):
\begin{equation}
\textrm{flux}=(5.3\times HR1+8.31)\times 10^{-12} \times
counts~s^{-1} \lbrack erg ~ cm^{-2}s^{-1}\rbrack
\end{equation}
the ROSAT count-rate of 0.111$\pm0.016$ c/s and hardness ratio
HR1=0.07$\pm0.13$ of 1RXS J201607.0+251645 yield a 0.1-2.4 keV flux
of $9.6\times10^{12} erg$ $cm^{2}$ $s^{-1}$.
Its distance is $19.9^{~+292.6}_{~-9.6}$~ pc, corresponding to the
parallax $\pi$ = 50.30 $\pm47.10$ mas \citep{ESA97}.
Hence $\textrm{log} L_{X} = 28.7^{~+2.4}_{~-0.6}~erg~s^{-1}$.
Combining its apparent magnitude of $10\fm6$,
we got $\textrm{log} L_{X}/L_{bol} \sim -3.27$.

Since the source is locating outside nearby molecular cloud,
interstellar absorption should be marginal. As for the internal
absorption, it's hard to be estimated \citep{Feigelson02,
Stassun04}.

Our calculated log$L_{X}$ is a little lower comparing to
statistic values of near-contact binaries made by Shaw et al. (1996),
while log$L_{X}$ is in the range of 29.09 to 30.55 ergs s$^{-1}$.
However, it is still consistent with their study, if taking into
account the large uncertainty of distance.
In fact, the value of log $L_{X}/L_{bol}$ is more reliable, since it is
independent of distance.
It is falling in the high end of statistic
range of -3.2 to -4.1, suggesting the binary is X-ray luminous.

The X-ray emission in the Algol systems are generally considered to
originate from chromospheric activity, interacting material, hot ($>10^{6}\mathrm{K}$)
disk \citep{Richards93, Retter05} or extended "halo" that pervades the entire system
\citep{Siarkowski96}.
Chromospheric activity is the most possible mechanism for our target.
Since both components of our target are cool stars with effective
temperature lower than that of the Sun, the surfaces on the two components
should be in deep convective. It is generally believed that convective
motions generate acoustic and MHD waves that propagate upward and heat
the chromosphere and corona \citep{Sarna98}, which can account for
the strong X-ray emission in this system. If true, thus its variability
outside eclipse could be easily understood, because strongly convective
envelope of cool stars can also account for comparative amount of dark spots.
Besides, the X-ray emission might also attribute to the
the interaction between infalling stream and the star \citep{Sarna98} or
extended "halo" that pervades the entire system \citep{Siarkowski96}.
We do not exclude their possibilities in our target, for the
near-contact configuration of the system, as discussed in later section.
For the mechanism of hot disk, spectral observation is needed to
identify its existence or absence.

\subsection{The evolutionary status}

The photometric solution suggests that 1RXS J201607.0+251645 is a
semi-detached binary, with the less massive component already
filling up its Roche lobe. For the primary, the size of its Roche
lobe can be determined according to the formula \citep{Eggleton83}
\begin{equation}
R_{L}^{i}=\frac{0.49(m_{i}/m_{j})^{\frac{2}{3}}}
{0.6(m_{i}/m_{j})^{\frac{2}{3}}+\ln{(1+(m_{i}/m_{j})^{\frac{1}{3}})}}
\end{equation}
And the fill-out factor is defined as f = r/R \citep{Zhai89}, where
$r$ denotes the mean relative radius. With the above two relations,
the filling factor of the primary is calculated to be 93.4\%. This
implies that the primary star is very close to fill up its Roche
lobe. Therefore, we may infer that the two components of 1RXS
J201607.0+251645 is in near-contact \citep{Shaw94}.

Suppose the primary is a main-sequence star, its mass and radius
could be given as $M_1 = 0.92M_{\odot}$, $R_1 = 0.92R_{\odot}$
\citep{Cox2000}, according to its spectral type G5. The
secondary component should be a K5 type star, according to its
temperature given by the photometric solution. Meanwhile, with mass
ratio and radius ratio, we could derive the absolute parameters of
the secondary as $M_2 = 0.82M_{\odot}$, $R_2 = 0.98R_{\odot}$. This
calculation gives an oversize configuration to the secondary
component.

Statistics show that semi-detached short-period (P$<$5d) Algols
should have lost significant amount of angular momentum
\citep{Ibanoglu06}. If angular momentum loss is continuing in this
binary, we could expect that its orbital period is decreasing (Yang
\& Wei 2009). With the on-going mass transfer, 1RXS J201607.0+251645
may be in thermal oscillation predicted by thermal relaxation
oscillation theory \citep{Lucy79}. Another possibility
is that it will fill up its primary component and finally evolve into a contact
binary \citep{Bradstreet94}.

\section{conclusion}

1RXS J201607.0+251645 is identified to an $EB$ type eclipsing binary
based on the morphology of its light curve. The photometric result
reveals that it is a semi-contact Algol type system with the less
massive component completely filling its Roche lobe. The photometric
 mass ratio and the orbital period are $0.895(\pm0.006)$ and $0\fd38805$,
 respectively.
Optical variability outside eclipse of the system showed varying
O'Connell effect in its phase folded light diagrams, which is likely
due to hot spots generated by accretion or dark spots from
chromospheric activity. The system is X-ray luminous with
$\textrm{log} L_{X}/L_{bol} \sim -3.27$. The actual process that
account for its X-ray emission is not clear yet, nevertheless,
possible mechanisms are discussed. The system is likely to comprise
a G5 primary and an oversize K5 secondary. It may be in thermal
oscillation or may evolve into a contact binary.

\begin{acknowledgements}
We wish to thank the referee for suggesting a number of improvements
to this paper. We are indebt to Dr. Yu-Lei QIU and Dr. Li CAO for
providing original version of reduction programs.
Also, we are grateful to night assistants, technical supports for their assistant of the observation, to the astronomers
in the Xinglong Observatory especially Dr. Jing-Yao HU, Dr. Xiao-Jun JIANG, Dr.
Jing-Song DENG and Dr. Jing WANG for helpful discussion, and to Wei-Kang ZHENG,
 Xu-Hui HAN and Li-Ping XIN for catching errors in
our manuscript. This work was supported by the National Natural
Science Foundation of China through grant No. 10778707 and No. 10473013. We have made
use of the $ROSAT$ data Archive of the Max-Planck-Institut fur
extraterrestrische Physik at Garching, Germany, as well as the
SIMBAD database, operated at CDS, Strausbourg, France.
\end{acknowledgements}

\clearpage

\begin{table}
\begin{center}
\caption[]{Details of target star, comparison star and check star in
our observation.}\label{Tab1}
 \begin{tabular}{ccccc}
  \hline
star    & $\alpha_{2000.0}$  & $\delta_{2000.0}$   &  Mag($V$) \\
\hline
target  & 20:16:06.98  & 25:16:53.8 & $10\fm6$ \\
C1      & 20:16:20.02  & 25:20:13.7 & $10\fm3$  \\
C2      & 20:16:06.90  & 25:19:21.8 & $11\fm0$  \\
\hline
\end{tabular}
\end{center}
\end{table}

\begin{table}
\begin{center}
\caption[]{Light minimum times of 1RXS
J201607.0+251645.}\label{Tab2}
 \begin{tabular}{ccc}
  \hline
HJD & Band & Error \\
\hline
2454770.6629    &R  &0.0001  \\
2454771.6314    &VR &0.0002  \\
2454774.7470    &VR &0.0003  \\
2454775.7047    &VR &0.0007  \\
2454776.6755    &R  &0.0004  \\
\hline
\end{tabular}
\end{center}
\end{table}

\begin{table}
\caption{Light-curve solution for 1RXS J201607.0+251645.}
\label{Tab3} \centering
\begin{tabular}{lc}
\hline
Parameters               &Values \\
\hline
$i(^{\circ})$            &$70.16(\pm0.06)$ \\
$q=M_{2}/M_{1}$          &$0.895(\pm0.006)$\\
$T_{1}(K)$               &5560             \\
$T_{2}(K)$               &$4301(\pm8)$    \\
$\Omega_{1}$             &$3.7364(\pm0.0118)$  \\
$\Omega_{2}=\Omega_{in}$ &$3.5856$  \\
$L_{1}/(L_{1}+L_{2})_{V}$&$0.8298(\pm0.0070)$  \\
$L_{1}/(L_{1}+L_{2})_{R}$&$0.7717(\pm0.0056)$  \\
$x_{2V}$                 &0.85          \\
$x_{2R}$                 &0.68          \\
$r_{1} (pole)$  &$0.3454(\pm0.0015)$\\
$r_{1} (side)$  &$0.3602(\pm0.0018)$\\
$r_{1} (back)$  &$0.3826(\pm0.0024)$\\
$r_{1} (point)$ &$0.4116(\pm0.0041)$\\
$r_{2} (pole)$  &$0.3465(\pm0.0005)$\\
$r_{2} (side)$  &$0.3635(\pm0.0006)$\\
$r_{2} (back)$  &$0.3948(\pm0.0006)$\\
$r_{2} (point)$ &$0.4881(\pm0.0022)$\\
$\sum(O-C)^{2}_{i}$ &$0.1828$       \\
\hline
\end{tabular}
\end{table}

\begin{figure}
\begin{center}
\includegraphics[width=8cm]{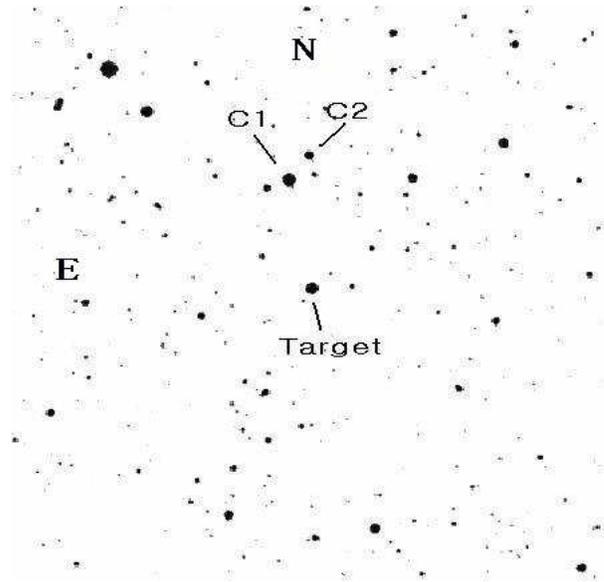}
\caption{Finding chart of 1RXS J201607.0+251645 made from a $V$ band
$CCD$ image taken at the 0.8-m telescope. The field of view is
$11'\times11'$. $C$1 and $C$2 are used as comparison and check
stars.} \label{Fig1}
\end{center}
\end{figure}

\begin{figure}
\begin{center}
\includegraphics[width=12cm]{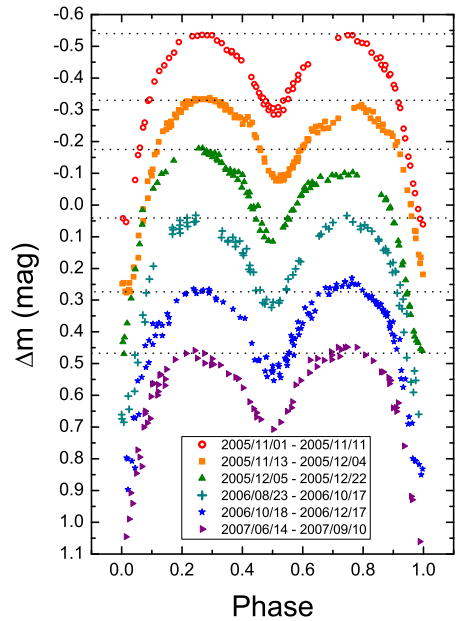}
\caption{Phase folded diagrams of 1RXS J201607.0+251645 from 2005 to 2007,
which were made with the data obtained in the 0.6-m telescope. They are
arbitrarily shifted for display purposes.} \label{Fig2}
\end{center}
\end{figure}

\begin{figure}
\begin{center}
\includegraphics[width=14cm]{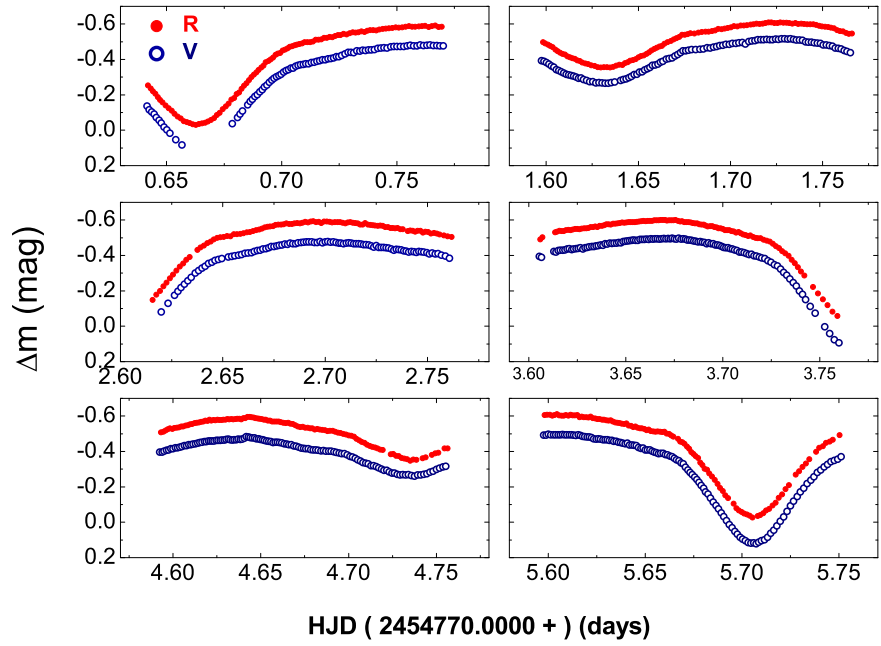}
\caption{$VR$-band light curves of 1RXS J201607.0+251645 covering 6
nights with the 0.8-m telescope.} \label{Fig4}
\end{center}
\end{figure}

\begin{figure}
\begin{center}
\includegraphics[width=6cm]{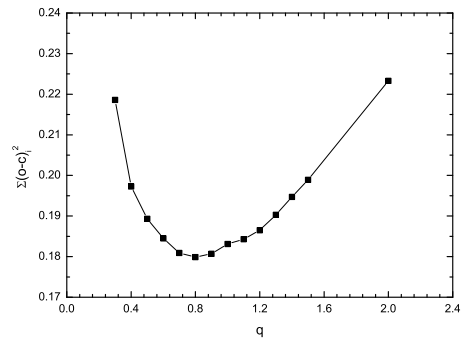}
\caption{The relation of $\sum-q$ for 1RXS J201607.0+251645.}
\label{Fig5}
\end{center}
\end{figure}

\begin{figure}
\begin{center}
\includegraphics[width=10cm]{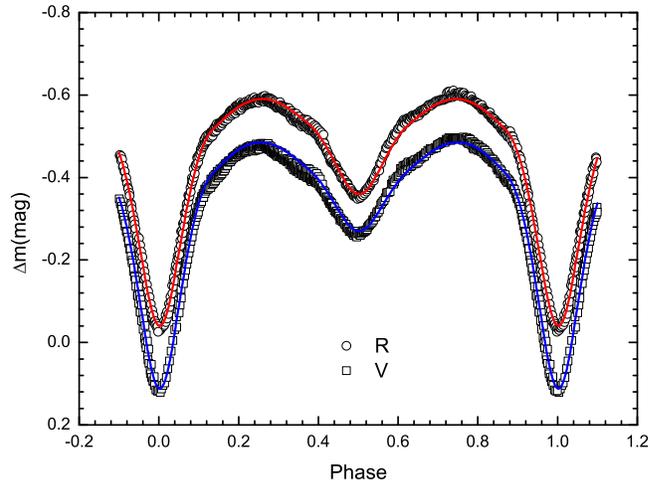}
\caption{Light curves of 1RXS J201607.0+251645. The open circles and
squares represent $V$ and $R$ observations. The red and blue lines
were calculated from the photometric solution.} \label{Fig6}
\end{center}
\end{figure}

\begin{figure}
\begin{center}
\includegraphics[width=11cm]{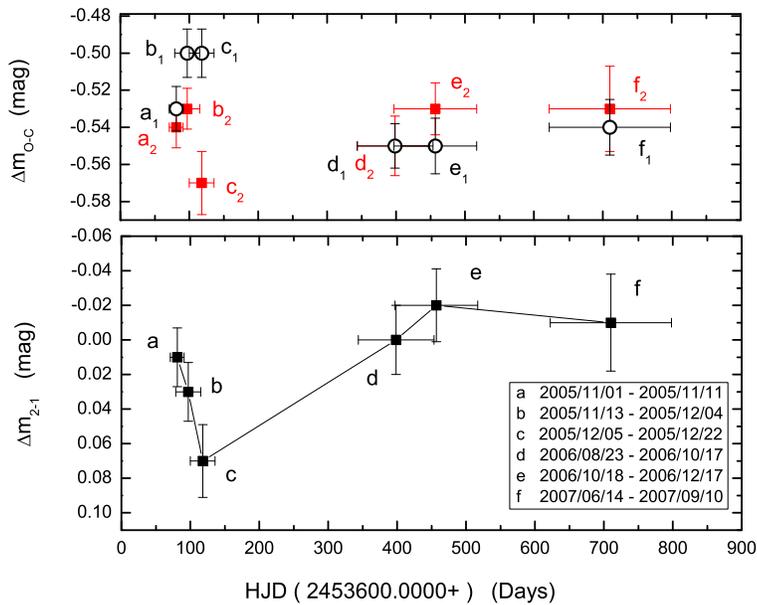}
\caption{The top panel: maximum brightness VS. HJD. Subscripts "1" and
"2" denote the primary maximum brightness($Max. I$, in magnitude) and
the secondary maximum brightness($Max. II$, in magnitude)).
The bottom panel: $Max.II-Max.I$ versus HJD.}
\label{Fig7}
\end{center}
\end{figure}

\begin{figure}
\begin{center}
\includegraphics[width=8cm]{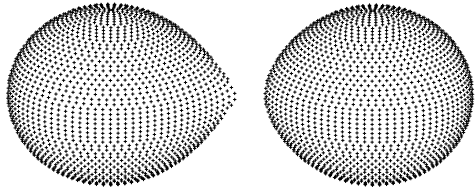}
\caption{3D representation for 1RXS J201607.0+251645 at phase 0.75.}
\label{Fig7}
\end{center}
\end{figure}

\clearpage
\appendix I
\begin{longtable}{ccccccccccc}
\caption[]{$V$ band observations of 1RXS
J201607.0+251645.}\\
\endfirsthead
\caption[]{(continued)}\\
\hline\noalign{\smallskip}
HJD &  & HJD &&  HJD & &  HJD & &  HJD & \\
+2454770 & $\Delta$(m) & +2454770 & $\Delta$(m) & +2454770 & $\Delta$(m) & +2454770 & $\Delta$(m) & +2454770 & $\Delta$(m)\\
\hline\noalign{\smallskip}
\endhead
  \hline\noalign{\smallskip}
HJD &  & HJD &&  HJD & &  HJD & &  HJD & \\
+2454770 & $\Delta$(m) & +2454770 & $\Delta$(m) & +2454770 & $\Delta$(m) & +2454770 & $\Delta$(m) & +2454770 & $\Delta$(m)\\
  \hline\noalign{\smallskip}
0.6415  &  -0.136  &  0.6425  &  -0.116  &  0.6436  &  -0.104  &   0.6446  &  -0.091  &  0.6456  &  -0.072  \\
0.6467  &  -0.057  &  0.6477  &  -0.041  &  0.6489  &  -0.018  &   0.6501  &  -0.003  &  0.6514  &  0.016   \\
0.6540  &  0.053   &  0.6567  &  0.083   &  0.6786  &  -0.037  &   0.6808  &  -0.072  &  0.6819  &  -0.091  \\
0.6830  &  -0.108  &  0.6852  &  -0.143  &  0.6863  &  -0.162  &   0.6873  &  -0.175  &  0.6884  &  -0.19   \\
0.6894  &  -0.202  &  0.6905  &  -0.217  &  0.6915  &  -0.229  &   0.6925  &  -0.245  &  0.6936  &  -0.256  \\
0.6946  &  -0.271  &  0.6957  &  -0.284  &  0.6967  &  -0.293  &   0.6977  &  -0.303  &  0.6988  &  -0.31   \\
0.6998  &  -0.316  &  0.7008  &  -0.329  &  0.7019  &  -0.338  &   0.7029  &  -0.342  &  0.7040  &  -0.351  \\
0.7050  &  -0.352  &  0.7060  &  -0.361  &  0.7071  &  -0.367  &   0.7082  &  -0.371  &  0.7094  &  -0.372  \\
0.7105  &  -0.376  &  0.7117  &  -0.376  &  0.7129  &  -0.382  &   0.7141  &  -0.385  &  0.7153  &  -0.389  \\
0.7165  &  -0.393  &  0.7177  &  -0.399  &  0.7189  &  -0.399  &   0.7201  &  -0.404  &  0.7213  &  -0.406  \\
0.7225  &  -0.409  &  0.7237  &  -0.416  &  0.7248  &  -0.416  &   0.7260  &  -0.422  &  0.7275  &  -0.431  \\
0.7288  &  -0.44   &  0.7300  &  -0.44   &  0.7313  &  -0.439  &   0.7326  &  -0.448  &  0.7339  &  -0.447  \\
0.7351  &  -0.453  &  0.7364  &  -0.45   &  0.7377  &  -0.454  &   0.7389  &  -0.454  &  0.7403  &  -0.456  \\
0.7417  &  -0.463  &  0.7430  &  -0.465  &  0.7444  &  -0.467  &   0.7458  &  -0.465  &  0.7473  &  -0.471  \\
0.7487  &  -0.473  &  0.7501  &  -0.474  &  0.7515  &  -0.473  &   0.7530  &  -0.473  &  0.7544  &  -0.476  \\
0.7558  &  -0.481  &  0.7573  &  -0.477  &  0.7587  &  -0.476  &   0.7601  &  -0.482  &  0.7616  &  -0.479  \\
0.7630  &  -0.481  &  0.7644  &  -0.482  &  0.7659  &  -0.479  &   0.7673  &  -0.479  &  0.7687  &  -0.477  \\
0.7702  &  -0.476  &  1.5975  &  -0.393  &  1.5988  &  -0.389  &   1.6001  &  -0.383  &  1.6015  &  -0.373  \\
1.6028  &  -0.364  &  1.6041  &  -0.356  &  1.6055  &  -0.343  &   1.6068  &  -0.338  &  1.6082  &  -0.331  \\
1.6095  &  -0.324  &  1.6108  &  -0.316  &  1.6122  &  -0.313  &   1.6135  &  -0.31   &  1.6148  &  -0.305  \\
1.6162  &  -0.3    &  1.6175  &  -0.294  &  1.6189  &  -0.295  &   1.6202  &  -0.286  &  1.6219  &  -0.282  \\
1.6236  &  -0.274  &  1.6253  &  -0.275  &  1.6269  &  -0.268  &   1.6286  &  -0.269  &  1.6303  &  -0.267  \\
1.6320  &  -0.266  &  1.6337  &  -0.266  &  1.6354  &  -0.27   &   1.6371  &  -0.272  &  1.6404  &  -0.281  \\
1.6419  &  -0.287  &  1.6433  &  -0.29   &  1.6448  &  -0.297  &   1.6462  &  -0.304  &  1.6477  &  -0.314  \\
1.6492  &  -0.316  &  1.6506  &  -0.329  &  1.6521  &  -0.331  &   1.6535  &  -0.341  &  1.6549  &  -0.347  \\
1.6563  &  -0.355  &  1.6577  &  -0.366  &  1.6592  &  -0.369  &   1.6606  &  -0.374  &  1.6620  &  -0.385  \\
1.6634  &  -0.39   &  1.6651  &  -0.401  &  1.6667  &  -0.405  &   1.6684  &  -0.409  &  1.6698  &  -0.419  \\
1.6713  &  -0.428  &  1.6728  &  -0.439  &  1.6743  &  -0.448  &   1.6758  &  -0.449  &  1.6773  &  -0.453  \\
1.6789  &  -0.455  &  1.6811  &  -0.455  &  1.6834  &  -0.458  &   1.6853  &  -0.46   &  1.6873  &  -0.466  \\
1.6890  &  -0.464  &  1.6908  &  -0.469  &  1.6925  &  -0.473  &   1.6942  &  -0.479  &  1.6959  &  -0.48   \\
1.6976  &  -0.486  &  1.6993  &  -0.486  &  1.7010  &  -0.49   &   1.7028  &  -0.49   &  1.7046  &  -0.494  \\
1.7064  &  -0.496  &  1.7081  &  -0.488  &  1.7100  &  -0.501  &   1.7116  &  -0.505  &  1.7132  &  -0.506  \\
1.7148  &  -0.511  &  1.7164  &  -0.512  &  1.7180  &  -0.511  &   1.7197  &  -0.514  &  1.7213  &  -0.512  \\
1.7229  &  -0.511  &  1.7245  &  -0.515  &  1.7261  &  -0.513  &   1.7277  &  -0.517  &  1.7294  &  -0.516  \\
1.7310  &  -0.515  &  1.7326  &  -0.516  &  1.7342  &  -0.513  &   1.7358  &  -0.51   &  1.7374  &  -0.51   \\
1.7391  &  -0.508  &  1.7407  &  -0.5    &  1.7423  &  -0.497  &   1.7439  &  -0.497  &  1.7455  &  -0.495  \\
1.7471  &  -0.493  &  1.7488  &  -0.49   &  1.7504  &  -0.482  &   1.7520  &  -0.482  &  1.7536  &  -0.479  \\
1.7553  &  -0.475  &  1.7569  &  -0.465  &  1.7585  &  -0.463  &   1.7601  &  -0.46   &  1.7617  &  -0.45   \\
1.7633  &  -0.445  &  1.7650  &  -0.439  &  2.6199  &  -0.081  &   2.6232  &  -0.129  &  2.6264  &  -0.175  \\
2.6280  &  -0.198  &  2.6296  &  -0.218  &  2.6312  &  -0.237  &   2.6329  &  -0.257  &  2.6345  &  -0.276  \\
2.6365  &  -0.3    &  2.6382  &  -0.315  &  2.6398  &  -0.328  &   2.6414  &  -0.339  &  2.6430  &  -0.354  \\
2.6446  &  -0.361  &  2.6462  &  -0.372  &  2.6479  &  -0.378  &   2.6495  &  -0.382  &  2.6527  &  -0.392  \\
2.6543  &  -0.395  &  2.6559  &  -0.399  &  2.6576  &  -0.402  &   2.6592  &  -0.405  &  2.6608  &  -0.413  \\
2.6623  &  -0.414  &  2.6639  &  -0.415  &  2.6654  &  -0.422  &   2.6668  &  -0.428  &  2.6682  &  -0.429  \\
2.6696  &  -0.437  &  2.6710  &  -0.439  &  2.6724  &  -0.443  &   2.6737  &  -0.45   &  2.6750  &  -0.451  \\
2.6763  &  -0.454  &  2.6776  &  -0.458  &  2.6789  &  -0.46   &   2.6803  &  -0.464  &  2.6816  &  -0.463  \\
2.6829  &  -0.469  &  2.6842  &  -0.466  &  2.6855  &  -0.466  &   2.6868  &  -0.472  &  2.6882  &  -0.474  \\
2.6895  &  -0.475  &  2.6908  &  -0.477  &  2.6921  &  -0.476  &   2.6934  &  -0.476  &  2.6947  &  -0.473  \\
2.6961  &  -0.473  &  2.6974  &  -0.479  &  2.6987  &  -0.472  &   2.7000  &  -0.476  &  2.7013  &  -0.479  \\
2.7026  &  -0.475  &  2.7040  &  -0.477  &  2.7053  &  -0.473  &   2.7066  &  -0.47   &  2.7078  &  -0.469  \\
2.7090  &  -0.467  &  2.7103  &  -0.469  &  2.7115  &  -0.471  &   2.7127  &  -0.469  &  2.7139  &  -0.466  \\
2.7151  &  -0.466  &  2.7164  &  -0.466  &  2.7176  &  -0.464  &   2.7188  &  -0.461  &  2.7200  &  -0.461  \\
2.7213  &  -0.46   &  2.7225  &  -0.454  &  2.7237  &  -0.447  &   2.7249  &  -0.455  &  2.7262  &  -0.446  \\
2.7274  &  -0.446  &  2.7286  &  -0.445  &  2.7298  &  -0.444  &   2.7310  &  -0.441  &  2.7323  &  -0.436  \\
2.7335  &  -0.43   &  2.7347  &  -0.435  &  2.7359  &  -0.434  &   2.7372  &  -0.431  &  2.7384  &  -0.427  \\
2.7396  &  -0.425  &  2.7408  &  -0.42   &  2.7421  &  -0.422  &   2.7433  &  -0.421  &  2.7445  &  -0.418  \\
2.7457  &  -0.422  &  2.7469  &  -0.421  &  2.7482  &  -0.42   &   2.7496  &  -0.415  &  2.7509  &  -0.408  \\
2.7523  &  -0.414  &  2.7537  &  -0.408  &  2.7551  &  -0.41   &   2.7568  &  -0.402  &  2.7587  &  -0.396  \\
2.7607  &  -0.384  &  3.6054  &  -0.394  &  3.6065  &  -0.389  &   3.6129  &  -0.422  &  3.6140  &  -0.42   \\
3.6152  &  -0.428  &  3.6163  &  -0.429  &  3.6175  &  -0.43   &   3.6186  &  -0.431  &  3.6198  &  -0.436  \\
3.6209  &  -0.438  &  3.6221  &  -0.434  &  3.6232  &  -0.437  &   3.6244  &  -0.439  &  3.6255  &  -0.44   \\
3.6267  &  -0.443  &  3.6279  &  -0.437  &  3.6290  &  -0.444  &   3.6302  &  -0.446  &  3.6313  &  -0.448  \\
3.6325  &  -0.451  &  3.6336  &  -0.451  &  3.6348  &  -0.454  &   3.6359  &  -0.455  &  3.6371  &  -0.46   \\
3.6382  &  -0.458  &  3.6394  &  -0.466  &  3.6405  &  -0.465  &   3.6417  &  -0.47   &  3.6429  &  -0.468  \\
3.6440  &  -0.47   &  3.6452  &  -0.47   &  3.6463  &  -0.473  &   3.6475  &  -0.472  &  3.6486  &  -0.479  \\
3.6497  &  -0.48   &  3.6508  &  -0.482  &  3.6519  &  -0.483  &   3.6530  &  -0.485  &  3.6541  &  -0.487  \\
3.6552  &  -0.486  &  3.6563  &  -0.485  &  3.6574  &  -0.49   &   3.6585  &  -0.491  &  3.6596  &  -0.491  \\
3.6607  &  -0.493  &  3.6619  &  -0.49   &  3.6630  &  -0.494  &   3.6641  &  -0.493  &  3.6652  &  -0.495  \\
3.6663  &  -0.494  &  3.6674  &  -0.493  &  3.6685  &  -0.495  &   3.6696  &  -0.496  &  3.6707  &  -0.495  \\
3.6718  &  -0.496  &  3.6729  &  -0.495  &  3.6740  &  -0.498  &   3.6751  &  -0.495  &  3.6762  &  -0.493  \\
3.6773  &  -0.497  &  3.6785  &  -0.492  &  3.6796  &  -0.494  &   3.6807  &  -0.492  &  3.6818  &  -0.487  \\
3.6829  &  -0.487  &  3.6840  &  -0.486  &  3.6851  &  -0.482  &   3.6862  &  -0.48   &  3.6873  &  -0.476  \\
3.6884  &  -0.479  &  3.6895  &  -0.475  &  3.6906  &  -0.473  &   3.6917  &  -0.467  &  3.6928  &  -0.471  \\
3.6940  &  -0.467  &  3.6951  &  -0.466  &  3.6962  &  -0.46   &   3.6973  &  -0.46   &  3.6984  &  -0.455  \\
3.6995  &  -0.45   &  3.7006  &  -0.445  &  3.7017  &  -0.444  &   3.7028  &  -0.444  &  3.7039  &  -0.439  \\
3.7050  &  -0.437  &  3.7061  &  -0.432  &  3.7073  &  -0.428  &   3.7086  &  -0.42   &  3.7098  &  -0.42   \\
3.7111  &  -0.416  &  3.7123  &  -0.41   &  3.7136  &  -0.405  &   3.7149  &  -0.407  &  3.7161  &  -0.399  \\
3.7174  &  -0.396  &  3.7188  &  -0.392  &  3.7201  &  -0.386  &   3.7217  &  -0.383  &  3.7233  &  -0.37   \\
3.7249  &  -0.361  &  3.7266  &  -0.348  &  3.7282  &  -0.341  &   3.7298  &  -0.323  &  3.7314  &  -0.304  \\
3.7330  &  -0.288  &  3.7346  &  -0.268  &  3.7363  &  -0.247  &   3.7379  &  -0.227  &  3.7395  &  -0.199  \\
3.7411  &  -0.181  &  3.7428  &  -0.151  &  3.7451  &  -0.113  &   3.7476  &  -0.073  &  3.7526  &  0.003   \\
3.7551  &  0.042   &  3.7576  &  0.076   &  3.7600  &  0.093   &   4.5924  &  -0.396  &  4.5937  &  -0.396  \\
4.5949  &  -0.402  &  4.5962  &  -0.407  &  4.5975  &  -0.408  &   4.5987  &  -0.415  &  4.6000  &  -0.415  \\
4.6013  &  -0.421  &  4.6026  &  -0.421  &  4.6038  &  -0.424  &   4.6051  &  -0.428  &  4.6064  &  -0.433  \\
4.6076  &  -0.434  &  4.6089  &  -0.438  &  4.6102  &  -0.438  &   4.6114  &  -0.441  &  4.6127  &  -0.443  \\
4.6140  &  -0.447  &  4.6153  &  -0.45   &  4.6165  &  -0.453  &   4.6178  &  -0.455  &  4.6189  &  -0.456  \\
4.6201  &  -0.459  &  4.6212  &  -0.459  &  4.6223  &  -0.459  &   4.6234  &  -0.46   &  4.6246  &  -0.462  \\
4.6257  &  -0.463  &  4.6268  &  -0.462  &  4.6280  &  -0.464  &   4.6291  &  -0.465  &  4.6302  &  -0.468  \\
4.6314  &  -0.47   &  4.6325  &  -0.466  &  4.6337  &  -0.464  &   4.6348  &  -0.467  &  4.6360  &  -0.466  \\
4.6371  &  -0.469  &  4.6383  &  -0.468  &  4.6394  &  -0.469  &   4.6406  &  -0.475  &  4.6417  &  -0.483  \\
4.6429  &  -0.482  &  4.6440  &  -0.479  &  4.6452  &  -0.476  &   4.6464  &  -0.476  &  4.6475  &  -0.472  \\
4.6487  &  -0.471  &  4.6498  &  -0.468  &  4.6510  &  -0.466  &   4.6521  &  -0.463  &  4.6533  &  -0.46   \\
4.6545  &  -0.458  &  4.6557  &  -0.455  &  4.6570  &  -0.451  &   4.6582  &  -0.452  &  4.6595  &  -0.452  \\
4.6607  &  -0.45   &  4.6619  &  -0.448  &  4.6633  &  -0.443  &   4.6647  &  -0.442  &  4.6660  &  -0.439  \\
4.6674  &  -0.439  &  4.6688  &  -0.432  &  4.6702  &  -0.43   &   4.6716  &  -0.425  &  4.6730  &  -0.419  \\
4.6743  &  -0.418  &  4.6757  &  -0.415  &  4.6771  &  -0.415  &   4.6785  &  -0.41   &  4.6799  &  -0.409  \\
4.6813  &  -0.409  &  4.6827  &  -0.407  &  4.6842  &  -0.403  &   4.6856  &  -0.405  &  4.6871  &  -0.402  \\
4.6886  &  -0.402  &  4.6901  &  -0.401  &  4.6916  &  -0.399  &   4.6931  &  -0.395  &  4.6946  &  -0.391  \\
4.6961  &  -0.388  &  4.6976  &  -0.388  &  4.6992  &  -0.381  &   4.7007  &  -0.372  &  4.7022  &  -0.368  \\
4.7037  &  -0.362  &  4.7052  &  -0.353  &  4.7067  &  -0.346  &   4.7082  &  -0.337  &  4.7097  &  -0.333  \\
4.7112  &  -0.33   &  4.7127  &  -0.32   &  4.7142  &  -0.319  &   4.7157  &  -0.314  &  4.7172  &  -0.307  \\
4.7187  &  -0.301  &  4.7202  &  -0.294  &  4.7217  &  -0.294  &   4.7232  &  -0.285  &  4.7247  &  -0.278  \\
4.7262  &  -0.278  &  4.7278  &  -0.271  &  4.7293  &  -0.268  &   4.7309  &  -0.269  &  4.7325  &  -0.267  \\
4.7341  &  -0.27   &  4.7358  &  -0.264  &  4.7376  &  -0.26   &   4.7394  &  -0.267  &  4.7411  &  -0.267  \\
4.7429  &  -0.271  &  4.7446  &  -0.274  &  4.7464  &  -0.284  &   4.7482  &  -0.289  &  4.7500  &  -0.298  \\
4.7519  &  -0.306  &  4.7537  &  -0.313  &  4.7554  &  -0.316  &   5.5976  &  -0.493  &  5.5989  &  -0.492  \\
5.6001  &  -0.497  &  5.6014  &  -0.495  &  5.6027  &  -0.494  &   5.6039  &  -0.495  &  5.6051  &  -0.495  \\
5.6062  &  -0.495  &  5.6074  &  -0.496  &  5.6085  &  -0.495  &   5.6097  &  -0.495  &  5.6109  &  -0.495  \\
5.6120  &  -0.494  &  5.6132  &  -0.492  &  5.6143  &  -0.49   &   5.6155  &  -0.494  &  5.6166  &  -0.49   \\
5.6178  &  -0.486  &  5.6189  &  -0.483  &  5.6201  &  -0.482  &   5.6212  &  -0.482  &  5.6224  &  -0.478  \\
5.6235  &  -0.471  &  5.6247  &  -0.469  &  5.6259  &  -0.472  &   5.6270  &  -0.47   &  5.6282  &  -0.466  \\
5.6293  &  -0.467  &  5.6305  &  -0.466  &  5.6316  &  -0.463  &   5.6328  &  -0.463  &  5.6339  &  -0.458  \\
5.6351  &  -0.457  &  5.6362  &  -0.453  &  5.6374  &  -0.45   &   5.6385  &  -0.447  &  5.6397  &  -0.444  \\
5.6408  &  -0.447  &  5.6420  &  -0.437  &  5.6432  &  -0.437  &   5.6443  &  -0.435  &  5.6455  &  -0.434  \\
5.6466  &  -0.423  &  5.6478  &  -0.418  &  5.6489  &  -0.419  &   5.6501  &  -0.409  &  5.6512  &  -0.409  \\
5.6524  &  -0.408  &  5.6535  &  -0.401  &  5.6547  &  -0.396  &   5.6558  &  -0.393  &  5.6570  &  -0.388  \\
5.6581  &  -0.389  &  5.6593  &  -0.39   &  5.6605  &  -0.382  &   5.6616  &  -0.38   &  5.6628  &  -0.374  \\
5.6639  &  -0.371  &  5.6651  &  -0.362  &  5.6662  &  -0.353  &   5.6674  &  -0.348  &  5.6685  &  -0.336  \\
5.6697  &  -0.33   &  5.6708  &  -0.311  &  5.6720  &  -0.296  &   5.6731  &  -0.282  &  5.6743  &  -0.27   \\
5.6755  &  -0.255  &  5.6766  &  -0.241  &  5.6778  &  -0.226  &   5.6789  &  -0.209  &  5.6801  &  -0.193  \\
5.6812  &  -0.172  &  5.6824  &  -0.159  &  5.6837  &  -0.136  &   5.6851  &  -0.114  &  5.6865  &  -0.095  \\
5.6879  &  -0.075  &  5.6893  &  -0.053  &  5.6906  &  -0.032  &   5.6920  &  -0.01   &  5.6934  &  0.013   \\
5.6948  &  0.03    &  5.6962  &  0.051   &  5.6976  &  0.068   &   5.6990  &  0.085   &  5.7003  &  0.095   \\
5.7017  &  0.104   &  5.7031  &  0.115   &  5.7045  &  0.118   &   5.7059  &  0.117   &  5.7073  &  0.121   \\
5.7086  &  0.113   &  5.7100  &  0.104   &  5.7115  &  0.098   &   5.7131  &  0.085   &  5.7147  &  0.063   \\
5.7163  &  0.045   &  5.7179  &  0.021   &  5.7196  &  -0.005  &   5.7214  &  -0.031  &  5.7231  &  -0.062  \\
5.7249  &  -0.093  &  5.7266  &  -0.119  &  5.7284  &  -0.146  &   5.7301  &  -0.171  &  5.7318  &  -0.193  \\
5.7336  &  -0.218  &  5.7354  &  -0.238  &  5.7371  &  -0.257  &   5.7388  &  -0.281  &  5.7406  &  -0.296  \\
5.7423  &  -0.313  &  5.7441  &  -0.327  &  5.7458  &  -0.342  &   5.7475  &  -0.349  &  5.7493  &  -0.357  \\
5.7510  &  -0.37   &  &       &  &       &  &       &  &       \\
  \noalign{\smallskip}\hline
\end{longtable}

\clearpage
\begin{longtable}{ccccccccccc}
\caption[]{$R$ band observations of 1RXS
J201607.0+251645.}\\
\endfirsthead
\caption[]{(continued)}\\
\hline\noalign{\smallskip}
HJD &  & HJD &&  HJD & &  HJD & &  HJD & \\
+2454770 & $\Delta$(m) & +2454770 & $\Delta$(m) & +2454770 & $\Delta$(m) & +2454770 & $\Delta$(m) & +2454770 & $\Delta$(m)\\
\hline\noalign{\smallskip}
\endhead
  \hline\noalign{\smallskip}

HJD     &          & HJD & & HJD & & HJD  & & HJD \\
+2454770 & $\Delta$(m) & +2454770 & $\Delta$(m) & +2454770 & $\Delta$(m) & +2454770 & $\Delta$(m) & +2454770 & $\Delta$(m)\\
  \hline\noalign{\smallskip}
0.6420  &  -0.253  &  0.6431  &  -0.236  &  0.6441  &  -0.22   &  0.6452  &   -0.207  &  0.6462  &  -0.187  \\
0.6472  &  -0.172  &  0.6483  &  -0.157  &  0.6496  &  -0.14   &  0.6509  &   -0.124  &  0.6522  &  -0.11   \\
0.6535  &  -0.093  &  0.6548  &  -0.078  &  0.6561  &  -0.066  &  0.6575  &   -0.052  &  0.6588  &  -0.046  \\
0.6601  &  -0.04   &  0.6614  &  -0.034  &  0.6627  &  -0.03   &  0.6641  &   -0.036  &  0.6654  &  -0.039  \\
0.6667  &  -0.042  &  0.6680  &  -0.049  &  0.6693  &  -0.058  &  0.6706  &   -0.069  &  0.6719  &  -0.086  \\
0.6731  &  -0.099  &  0.6744  &  -0.119  &  0.6757  &  -0.135  &  0.6769  &   -0.15   &  0.6782  &  -0.173  \\
0.6793  &  -0.182  &  0.6804  &  -0.205  &  0.6815  &  -0.213  &  0.6826  &   -0.235  &  0.6837  &  -0.252  \\
0.6847  &  -0.265  &  0.6858  &  -0.283  &  0.6869  &  -0.3    &  0.6880  &   -0.315  &  0.6890  &  -0.326  \\
0.6900  &  -0.341  &  0.6911  &  -0.354  &  0.6921  &  -0.365  &  0.6931  &   -0.384  &  0.6942  &  -0.391  \\
0.6952  &  -0.4    &  0.6963  &  -0.41   &  0.6973  &  -0.427  &  0.6983  &   -0.433  &  0.6994  &  -0.437  \\
0.7004  &  -0.45   &  0.7015  &  -0.456  &  0.7025  &  -0.463  &  0.7035  &   -0.472  &  0.7046  &  -0.475  \\
0.7056  &  -0.48   &  0.7066  &  -0.484  &  0.7077  &  -0.487  &  0.7088  &   -0.49   &  0.7100  &  -0.498  \\
0.7111  &  -0.497  &  0.7123  &  -0.498  &  0.7135  &  -0.501  &  0.7147  &   -0.511  &  0.7160  &  -0.511  \\
0.7172  &  -0.514  &  0.7184  &  -0.522  &  0.7196  &  -0.522  &  0.7208  &   -0.528  &  0.7220  &  -0.526  \\
0.7232  &  -0.532  &  0.7244  &  -0.537  &  0.7255  &  -0.538  &  0.7268  &   -0.544  &  0.7282  &  -0.542  \\
0.7295  &  -0.549  &  0.7308  &  -0.553  &  0.7320  &  -0.551  &  0.7333  &   -0.554  &  0.7346  &  -0.558  \\
0.7359  &  -0.556  &  0.7371  &  -0.562  &  0.7384  &  -0.565  &  0.7397  &   -0.566  &  0.7410  &  -0.564  \\
0.7424  &  -0.569  &  0.7438  &  -0.573  &  0.7452  &  -0.575  &  0.7466  &   -0.576  &  0.7480  &  -0.578  \\
0.7495  &  -0.582  &  0.7509  &  -0.579  &  0.7523  &  -0.582  &  0.7538  &   -0.585  &  0.7552  &  -0.588  \\
0.7566  &  -0.585  &  0.7580  &  -0.585  &  0.7595  &  -0.587  &  0.7609  &   -0.588  &  0.7623  &  -0.59   \\
0.7638  &  -0.584  &  0.7652  &  -0.589  &  0.7666  &  -0.591  &  0.7681  &   -0.583  &  0.7695  &  -0.585  \\
1.5982  &  -0.498  &  1.5995  &  -0.49   &  1.6009  &  -0.484  &  1.6022  &   -0.476  &  1.6035  &  -0.47   \\
1.6049  &  -0.455  &  1.6062  &  -0.447  &  1.6075  &  -0.438  &  1.6089  &   -0.432  &  1.6102  &  -0.428  \\
1.6116  &  -0.418  &  1.6129  &  -0.413  &  1.6142  &  -0.407  &  1.6156  &   -0.403  &  1.6169  &  -0.394  \\
1.6182  &  -0.391  &  1.6196  &  -0.382  &  1.6212  &  -0.378  &  1.6228  &   -0.374  &  1.6245  &  -0.367  \\
1.6262  &  -0.364  &  1.6279  &  -0.356  &  1.6296  &  -0.355  &  1.6313  &   -0.357  &  1.6330  &  -0.356  \\
1.6346  &  -0.355  &  1.6363  &  -0.359  &  1.6380  &  -0.365  &  1.6397  &   -0.37   &  1.6413  &  -0.368  \\
1.6427  &  -0.379  &  1.6442  &  -0.384  &  1.6456  &  -0.393  &  1.6471  &   -0.397  &  1.6485  &  -0.401  \\
1.6500  &  -0.413  &  1.6515  &  -0.42   &  1.6529  &  -0.423  &  1.6544  &   -0.432  &  1.6558  &  -0.44   \\
1.6572  &  -0.45   &  1.6586  &  -0.459  &  1.6600  &  -0.466  &  1.6614  &   -0.474  &  1.6628  &  -0.476  \\
1.6642  &  -0.489  &  1.6659  &  -0.498  &  1.6676  &  -0.504  &  1.6692  &   -0.513  &  1.6707  &  -0.521  \\
1.6722  &  -0.528  &  1.6737  &  -0.539  &  1.6752  &  -0.54   &  1.6767  &   -0.543  &  1.6782  &  -0.548  \\
1.6804  &  -0.546  &  1.6826  &  -0.552  &  1.6845  &  -0.555  &  1.6865  &   -0.557  &  1.6882  &  -0.557  \\
1.6900  &  -0.56   &  1.6917  &  -0.564  &  1.6934  &  -0.572  &  1.6952  &   -0.571  &  1.6969  &  -0.575  \\
1.6985  &  -0.578  &  1.7002  &  -0.582  &  1.7019  &  -0.589  &  1.7037  &   -0.587  &  1.7055  &  -0.592  \\
1.7073  &  -0.591  &  1.7090  &  -0.594  &  1.7109  &  -0.597  &  1.7125  &   -0.598  &  1.7141  &  -0.603  \\
1.7158  &  -0.601  &  1.7174  &  -0.603  &  1.7190  &  -0.605  &  1.7206  &   -0.61   &  1.7222  &  -0.609  \\
1.7238  &  -0.605  &  1.7255  &  -0.605  &  1.7271  &  -0.609  &  1.7287  &   -0.608  &  1.7303  &  -0.608  \\
1.7319  &  -0.605  &  1.7336  &  -0.603  &  1.7352  &  -0.605  &  1.7368  &   -0.6    &  1.7384  &  -0.602  \\
1.7400  &  -0.6    &  1.7416  &  -0.598  &  1.7433  &  -0.599  &  1.7449  &   -0.599  &  1.7465  &  -0.595  \\
1.7481  &  -0.59   &  1.7497  &  -0.587  &  1.7514  &  -0.585  &  1.7530  &   -0.583  &  1.7546  &  -0.579  \\
1.7562  &  -0.576  &  1.7578  &  -0.566  &  1.7594  &  -0.56   &  1.7611  &   -0.558  &  1.7627  &  -0.55   \\
1.7643  &  -0.544  &  1.7659  &  -0.546  &  2.6157  &  -0.149  &  2.6175  &   -0.178  &  2.6193  &  -0.2    \\
2.6209  &  -0.226  &  2.6225  &  -0.247  &  2.6241  &  -0.271  &  2.6257  &   -0.291  &  2.6273  &  -0.315  \\
2.6290  &  -0.338  &  2.6306  &  -0.355  &  2.6322  &  -0.375  &  2.6338  &   -0.392  &  2.6375  &  -0.431  \\
2.6391  &  -0.446  &  2.6407  &  -0.456  &  2.6424  &  -0.469  &  2.6440  &   -0.479  &  2.6456  &  -0.491  \\
2.6472  &  -0.499  &  2.6488  &  -0.503  &  2.6504  &  -0.504  &  2.6521  &   -0.51   &  2.6537  &  -0.509  \\
2.6553  &  -0.52   &  2.6569  &  -0.521  &  2.6585  &  -0.522  &  2.6601  &   -0.525  &  2.6617  &  -0.53   \\
2.6632  &  -0.532  &  2.6648  &  -0.538  &  2.6663  &  -0.543  &  2.6677  &   -0.547  &  2.6691  &  -0.551  \\
2.6704  &  -0.559  &  2.6718  &  -0.563  &  2.6731  &  -0.564  &  2.6745  &   -0.566  &  2.6758  &  -0.572  \\
2.6771  &  -0.575  &  2.6784  &  -0.575  &  2.6797  &  -0.577  &  2.6810  &   -0.582  &  2.6824  &  -0.578  \\
2.6837  &  -0.578  &  2.6850  &  -0.581  &  2.6863  &  -0.584  &  2.6876  &   -0.582  &  2.6889  &  -0.585  \\
2.6903  &  -0.59   &  2.6916  &  -0.587  &  2.6929  &  -0.591  &  2.6942  &   -0.594  &  2.6955  &  -0.589  \\
2.6968  &  -0.59   &  2.6981  &  -0.583  &  2.6995  &  -0.591  &  2.7008  &   -0.591  &  2.7021  &  -0.59   \\
2.7034  &  -0.587  &  2.7047  &  -0.588  &  2.7060  &  -0.588  &  2.7073  &   -0.589  &  2.7085  &  -0.585  \\
2.7098  &  -0.581  &  2.7110  &  -0.586  &  2.7122  &  -0.588  &  2.7134  &   -0.585  &  2.7147  &  -0.584  \\
2.7159  &  -0.581  &  2.7171  &  -0.575  &  2.7183  &  -0.575  &  2.7195  &   -0.583  &  2.7208  &  -0.575  \\
2.7220  &  -0.573  &  2.7232  &  -0.571  &  2.7244  &  -0.57   &  2.7257  &   -0.568  &  2.7269  &  -0.569  \\
2.7281  &  -0.568  &  2.7293  &  -0.561  &  2.7306  &  -0.561  &  2.7318  &   -0.556  &  2.7330  &  -0.556  \\
2.7342  &  -0.553  &  2.7354  &  -0.555  &  2.7367  &  -0.548  &  2.7379  &   -0.548  &  2.7403  &  -0.538  \\
2.7416  &  -0.542  &  2.7428  &  -0.54   &  2.7440  &  -0.534  &  2.7452  &   -0.54   &  2.7465  &  -0.539  \\
2.7477  &  -0.529  &  2.7489  &  -0.534  &  2.7503  &  -0.532  &  2.7517  &   -0.523  &  2.7531  &  -0.527  \\
2.7545  &  -0.524  &  2.7559  &  -0.519  &  2.7579  &  -0.514  &  2.7598  &   -0.509  &  2.7617  &  -0.504  \\
3.6060  &  -0.491  &  3.6071  &  -0.503  &  3.6135  &  -0.531  &  3.6146  &   -0.535  &  3.6158  &  -0.54   \\
3.6169  &  -0.538  &  3.6181  &  -0.538  &  3.6192  &  -0.543  &  3.6204  &   -0.544  &  3.6215  &  -0.543  \\
3.6227  &  -0.546  &  3.6239  &  -0.55   &  3.6250  &  -0.55   &  3.6262  &   -0.549  &  3.6273  &  -0.552  \\
3.6285  &  -0.554  &  3.6296  &  -0.554  &  3.6308  &  -0.556  &  3.6319  &   -0.561  &  3.6331  &  -0.56   \\
3.6342  &  -0.564  &  3.6354  &  -0.561  &  3.6365  &  -0.567  &  3.6377  &   -0.568  &  3.6389  &  -0.571  \\
3.6400  &  -0.576  &  3.6412  &  -0.576  &  3.6423  &  -0.574  &  3.6435  &   -0.579  &  3.6446  &  -0.579  \\
3.6458  &  -0.582  &  3.6469  &  -0.584  &  3.6481  &  -0.586  &  3.6492  &   -0.585  &  3.6503  &  -0.586  \\
3.6514  &  -0.588  &  3.6525  &  -0.589  &  3.6536  &  -0.588  &  3.6547  &   -0.588  &  3.6558  &  -0.59   \\
3.6569  &  -0.591  &  3.6580  &  -0.593  &  3.6591  &  -0.595  &  3.6603  &   -0.594  &  3.6614  &  -0.595  \\
3.6625  &  -0.596  &  3.6636  &  -0.595  &  3.6647  &  -0.599  &  3.6658  &   -0.598  &  3.6669  &  -0.598  \\
3.6680  &  -0.6    &  3.6691  &  -0.601  &  3.6702  &  -0.601  &  3.6713  &   -0.597  &  3.6724  &  -0.599  \\
3.6735  &  -0.599  &  3.6746  &  -0.598  &  3.6757  &  -0.6    &  3.6769  &   -0.595  &  3.6780  &  -0.592  \\
3.6791  &  -0.594  &  3.6802  &  -0.589  &  3.6813  &  -0.588  &  3.6824  &   -0.589  &  3.6835  &  -0.586  \\
3.6846  &  -0.584  &  3.6857  &  -0.581  &  3.6868  &  -0.582  &  3.6879  &   -0.576  &  3.6890  &  -0.575  \\
3.6901  &  -0.575  &  3.6912  &  -0.569  &  3.6923  &  -0.564  &  3.6935  &   -0.565  &  3.6946  &  -0.565  \\
3.6957  &  -0.561  &  3.6968  &  -0.559  &  3.6979  &  -0.553  &  3.6990  &   -0.554  &  3.7001  &  -0.55   \\
3.7012  &  -0.543  &  3.7023  &  -0.543  &  3.7034  &  -0.543  &  3.7045  &   -0.537  &  3.7056  &  -0.536  \\
3.7069  &  -0.529  &  3.7081  &  -0.532  &  3.7093  &  -0.525  &  3.7105  &   -0.518  &  3.7118  &  -0.515  \\
3.7131  &  -0.51   &  3.7143  &  -0.514  &  3.7156  &  -0.505  &  3.7169  &   -0.499  &  3.7181  &  -0.498  \\
3.7195  &  -0.497  &  3.7210  &  -0.489  &  3.7226  &  -0.483  &  3.7242  &   -0.478  &  3.7258  &  -0.467  \\
3.7275  &  -0.453  &  3.7291  &  -0.441  &  3.7307  &  -0.425  &  3.7323  &   -0.413  &  3.7339  &  -0.399  \\
3.7355  &  -0.376  &  3.7372  &  -0.356  &  3.7388  &  -0.334  &  3.7404  &   -0.309  &  3.7421  &  -0.287  \\
3.7466  &  -0.222  &  3.7491  &  -0.186  &  3.7516  &  -0.152  &  3.7541  &   -0.119  &  3.7566  &  -0.083  \\
3.7591  &  -0.058  &  4.5931  &  -0.509  &  4.5944  &  -0.511  &  4.5957  &   -0.519  &  4.5969  &  -0.525  \\
4.5982  &  -0.53   &  4.5995  &  -0.528  &  4.6007  &  -0.532  &  4.6020  &   -0.534  &  4.6033  &  -0.536  \\
4.6046  &  -0.541  &  4.6058  &  -0.542  &  4.6071  &  -0.546  &  4.6084  &   -0.548  &  4.6096  &  -0.553  \\
4.6109  &  -0.555  &  4.6122  &  -0.56   &  4.6134  &  -0.563  &  4.6147  &   -0.564  &  4.6160  &  -0.568  \\
4.6172  &  -0.57   &  4.6184  &  -0.574  &  4.6196  &  -0.572  &  4.6207  &   -0.579  &  4.6218  &  -0.574  \\
4.6230  &  -0.578  &  4.6241  &  -0.577  &  4.6252  &  -0.58   &  4.6264  &   -0.578  &  4.6275  &  -0.578  \\
4.6286  &  -0.579  &  4.6298  &  -0.58   &  4.6309  &  -0.583  &  4.6320  &   -0.582  &  4.6332  &  -0.584  \\
4.6343  &  -0.582  &  4.6355  &  -0.583  &  4.6366  &  -0.585  &  4.6378  &   -0.585  &  4.6389  &  -0.582  \\
4.6401  &  -0.588  &  4.6412  &  -0.589  &  4.6424  &  -0.596  &  4.6436  &   -0.595  &  4.6447  &  -0.594  \\
4.6459  &  -0.592  &  4.6470  &  -0.589  &  4.6482  &  -0.587  &  4.6493  &   -0.585  &  4.6505  &  -0.583  \\
4.6516  &  -0.579  &  4.6528  &  -0.58   &  4.6540  &  -0.575  &  4.6552  &   -0.575  &  4.6564  &  -0.569  \\
4.6577  &  -0.57   &  4.6589  &  -0.57   &  4.6602  &  -0.569  &  4.6614  &   -0.566  &  4.6627  &  -0.563  \\
4.6641  &  -0.56   &  4.6655  &  -0.562  &  4.6669  &  -0.557  &  4.6683  &   -0.555  &  4.6696  &  -0.549  \\
4.6710  &  -0.542  &  4.6724  &  -0.537  &  4.6738  &  -0.54   &  4.6752  &   -0.535  &  4.6766  &  -0.534  \\
4.6780  &  -0.53   &  4.6793  &  -0.53   &  4.6807  &  -0.524  &  4.6821  &   -0.527  &  4.6836  &  -0.524  \\
4.6850  &  -0.522  &  4.6865  &  -0.52   &  4.6879  &  -0.515  &  4.6894  &   -0.517  &  4.6909  &  -0.516  \\
4.6924  &  -0.511  &  4.6939  &  -0.506  &  4.6954  &  -0.504  &  4.6970  &   -0.499  &  4.6985  &  -0.496  \\
4.7000  &  -0.49   &  4.7015  &  -0.489  &  4.7030  &  -0.483  &  4.7045  &   -0.471  &  4.7060  &  -0.462  \\
4.7075  &  -0.454  &  4.7090  &  -0.448  &  4.7105  &  -0.438  &  4.7120  &   -0.432  &  4.7135  &  -0.427  \\
4.7150  &  -0.423  &  4.7165  &  -0.417  &  4.7180  &  -0.411  &  4.7195  &   -0.408  &  4.7240  &  -0.384  \\
4.7255  &  -0.386  &  4.7271  &  -0.375  &  4.7286  &  -0.373  &  4.7302  &   -0.364  &  4.7318  &  -0.36   \\
4.7334  &  -0.357  &  4.7351  &  -0.348  &  4.7369  &  -0.355  &  4.7386  &   -0.353  &  4.7421  &  -0.361  \\
4.7439  &  -0.366  &  4.7475  &  -0.381  &  4.7493  &  -0.389  &  4.7511  &   -0.393  &  4.7547  &  -0.418  \\
4.7564  &  -0.417  &  5.5982  &  -0.605  &  5.5995  &  -0.608  &  5.6008  &   -0.608  &  5.6021  &  -0.602  \\
5.6033  &  -0.605  &  5.6046  &  -0.611  &  5.6057  &  -0.604  &  5.6069  &   -0.602  &  5.6081  &  -0.602  \\
5.6092  &  -0.605  &  5.6104  &  -0.603  &  5.6115  &  -0.607  &  5.6127  &   -0.604  &  5.6138  &  -0.604  \\
5.6150  &  -0.605  &  5.6161  &  -0.594  &  5.6173  &  -0.597  &  5.6184  &   -0.594  &  5.6196  &  -0.594  \\
5.6207  &  -0.591  &  5.6219  &  -0.597  &  5.6230  &  -0.589  &  5.6242  &   -0.586  &  5.6254  &  -0.583  \\
5.6265  &  -0.583  &  5.6277  &  -0.58   &  5.6288  &  -0.58   &  5.6300  &   -0.576  &  5.6311  &  -0.574  \\
5.6323  &  -0.576  &  5.6334  &  -0.569  &  5.6346  &  -0.564  &  5.6357  &   -0.562  &  5.6369  &  -0.56   \\
5.6380  &  -0.557  &  5.6392  &  -0.555  &  5.6404  &  -0.558  &  5.6415  &   -0.547  &  5.6427  &  -0.545  \\
5.6438  &  -0.538  &  5.6450  &  -0.541  &  5.6461  &  -0.535  &  5.6473  &   -0.53   &  5.6484  &  -0.53   \\
5.6496  &  -0.528  &  5.6507  &  -0.521  &  5.6519  &  -0.523  &  5.6530  &   -0.515  &  5.6542  &  -0.514  \\
5.6553  &  -0.512  &  5.6565  &  -0.514  &  5.6577  &  -0.511  &  5.6588  &   -0.51   &  5.6600  &  -0.512  \\
5.6611  &  -0.501  &  5.6623  &  -0.499  &  5.6634  &  -0.491  &  5.6646  &   -0.483  &  5.6669  &  -0.472  \\
5.6680  &  -0.455  &  5.6692  &  -0.452  &  5.6703  &  -0.444  &  5.6715  &   -0.428  &  5.6727  &  -0.415  \\
5.6738  &  -0.397  &  5.6750  &  -0.386  &  5.6761  &  -0.373  &  5.6773  &   -0.363  &  5.6784  &  -0.338  \\
5.6796  &  -0.324  &  5.6807  &  -0.303  &  5.6819  &  -0.297  &  5.6831  &   -0.276  &  5.6845  &  -0.258  \\
5.6859  &  -0.24   &  5.6873  &  -0.216  &  5.6886  &  -0.194  &  5.6900  &   -0.177  &  5.6914  &  -0.153  \\
5.6928  &  -0.137  &  5.6956  &  -0.105  &  5.6970  &  -0.079  &  5.6983  &   -0.066  &  5.6997  &  -0.059  \\
5.7011  &  -0.049  &  5.7025  &  -0.044  &  5.7039  &  -0.037  &  5.7053  &   -0.026  &  5.7080  &  -0.036  \\
5.7094  &  -0.041  &  5.7109  &  -0.051  &  5.7124  &  -0.067  &  5.7140  &   -0.075  &  5.7157  &  -0.094  \\
5.7173  &  -0.111  &  5.7189  &  -0.138  &  5.7206  &  -0.158  &  5.7224  &   -0.188  &  5.7242  &  -0.21   \\
5.7276  &  -0.269  &  5.7294  &  -0.288  &  5.7311  &  -0.313  &  5.7329  &   -0.332  &  5.7346  &  -0.357  \\
5.7381  &  -0.399  &  5.7398  &  -0.421  &  5.7416  &  -0.429  &  5.7433  &   -0.449  &  5.7451  &  -0.456  \\
5.7468  &  -0.467  &  5.7503  &  -0.492  &  &       &  &       &  &       \\
  \noalign{\smallskip}\hline
\end{longtable}

\label{lastpage}
\end{document}